**Multilineage-differentiating stress-enduring cells alleviate neuropathic pain in mice by secreting TGF-β and IL-10**

Yayu Zhao[1†], Ying Fei[1†], Yunyun Cai[1], Zhongya Wei[1], Ying Chen[2], Yuhua Ji[1], Xue Chen[3], Dongmei Zhang[4*], Gang Chen [1,2,5*]

[1] Key Laboratory of Neuroregeneration of Jiangsu and the Ministry of Education, Co–innovation Center of Neuroregeneration, NMPA Key Laboratory for Research and Evaluation of Tissue Engineering Technology Products, Medical School of Nantong University, Nantong, Jiangsu Province, 226001, China

[2] Center for Basic Medical Research, Medical School of Nantong University, Nantong, Jiangsu Province, 226001, China

[3] Department of Histology and Embryology, Medical School of Jiangnan University, Wuxi, Jiangsu Province, 214122, China

[4] Medical Research Center, Affiliated Hospital 2 of Nantong University, Nantong 226001, Jiangsu Province, China

[5] Department of Anesthesiology, Affiliated Hospital of Nantong University, Nantong 226001, Jiangsu Province, China

*Corresponding author(s). E–mail(s): chengang6626@ntu.edu.cn; zdm_ntyy@163.com

†These authors contributed equally to this work.

**Abstract**

**Background:** Neuropathic pain is a chronic condition characterized by damage to and dysfunction of the peripheral or central nervous system. There are currently no effective treatment options available for neuropathic pain, and existing drugs often provide only temporary relief with potential side effects. Multilineage-differentiating stress-enduring (Muse) cells are characterized by high expansion potential, a stable phenotype and strong immunosuppression. These properties make them attractive candidates for therapeutics for neuropathic pain management. In this study, we conducted a series of experiments to evaluate the effect of Muse cells on neuropathic pain.

**Methods:** Human, rat, and mouse Muse cells were isolated and cultured via long-term digestion. After a single intrathecal injection, the analgesic effect of each agent was assessed by von–Frey and thermal stimulation of the soles of CCI mice. The differences between BMSCs and Muse cells were analyzed via protein profiling. Pain sensitivity in multiple mouse models of neuropathic pain was assessed by von–Frey and thermal radiation after a single intrathecal injection of Muse cells. Immunohistochemistry and qPCR were used to evaluate the activation and proliferation of glial cells in the spinal dorsal horn of the injured side of SNI model mice. To investigate the mechanisms underlying Muse cell migration, we performed immunohistochemistry, qPCR, and Transwell assays. Finally, the molecular mechanisms underlying the analgesic effects of Muse cells were investigated by ELISA and in vivo injection of neutralizing antibodies.

**Results:** Muse cells from different species demonstrated analgesic potential by reversing CCI-induced neuropathic pain. Protein profiling revealed a high degree of similarity between Muse cells and BMSCs. The intrathecal injection of Muse cells effectively reduced neuropathic pain in various mouse models, resulting in better analgesic effects than the administration of equivalent low doses of BMSCs. Immunohistochemical analysis and qPCR revealed the ability of Muse cells to inhibit spinal cord neuroinflammation caused by SNI. In addition, Transwell and ELISA revealed that Muse cells migrated through the injured dorsal root ganglion (DRG) via the CCR7–CCL21 chemotactic axis. In addition, the secretion of TGF-β and IL-10 by Muse cells was identified as the mechanism underlying the analgesic effect of Muse cells.

**Conclusions:** The capacity of Muse cells to mitigate neuroinflammation and produce analgesic effects via the modulation of TGF-β and IL-10 underscores their potential as promising therapeutic approaches for the treatment of neuropathic pain.

**Introduction**

Neuropathic pain caused by multiple insults to the nervous system is a refractory intractable pain[1]. Current treatments for neuropathic pain are insufficient. Bone marrow stromal cells (BMSCs), a population of progenitor cells that exist in adult bone marrow, have emerged as a major source for cell–based therapies because they are easy to collect and have high proliferation potential and strong immunosuppressive properties[2]. Mounting evidence suggests that the transplantation of BMSCs can achieve long-term relief of neuropathic pain[3,4], which provides a new method for the clinical treatment of neuropathic pain. Although the analgesic effects of BMSC transplantation on animal pain models are encouraging, there are still problems to overcome before their clinical application. The greatest barrier is the difficulty in establishing quality control standards for BMSCs that are suitable for clinical use.

BMSCs are a group of cells with nonuniform antigenic phenotypes, and their entire composition has not been fully identified[5]. Therefore, it is unclear which cell subsets of BMSCs have analgesic effects. During the process of culture and amplification, the proportions of different subsets of BMSCs always change. Because the BMSCs used in different laboratories may contain different proportions of different cell subsets, inconsistent research results on the analgesic effects of BMSCs in different laboratories may occur. Even in the same laboratory, BMSCs may undergo significant changes in cell characteristics, such as loss of multilineage differentiation and analgesic effects, after multiple passages.[6]. To date, no commercial BMSCs have been used to treat chronic pain in the clinic. At present, most clinical trials of BMSC transplantation have used acute isolation of autologous BMSCs or monocytes, and patients have to undergo considerable pain to extract enough bone marrow for transplantation[7]. In addition, BMSC drugs that include multiple cell subsets not only increase production and use costs but also may reduce analgesic effects due to the antagonistic effects of certain unknown cell subsets[8,9]. In summary, in a mixture of BMSCs, obtaining a single type of cell subset that is safe, stable, and durable is the key to promoting the clinical translation of the research of BMSCs in the treatment of neuropathic pain.

Multilineage-differentiating stress-enduring (Muse) cells, which were first described in 2010, constitute a small percentage of BMSCs, can generate cells representative of all three germ layers from a single cell, and they are nontumorigenic and self-renewable[10,11]. Muse cells are particularly unique compared with other stem cells in that they efficiently migrate and integrate into damaged tissue when supplied into the bloodstream and spontaneously differentiate into cells compatible with the homing tissue[12,13]. Therefore, current research on Muse cells has focused on their ability to differentiate and repair. Muse cells have emerged as a novel source for cell-based therapies, with a diverse spectrum of potential clinical applications. Due to their high expansion potential, genetic stability, stable phenotype, and strong immunosuppressive properties, Muse cells can be exploited for successful autologous and heterologous transplantation without the need of immune suppressants[10]. However, whether Muse cells, like BMSCs, play an analgesic role by secreting a variety of anti-inflammatory factors has not yet been reported. In this study, we investigated whether Muse cells can be used to treat neuropathic pain.

**Materials and methods**

The work has been reported in line with the ARRIVE guidelines 2.0.

**Animals and surgery.** Adult male ICR mice (25 ± 2 g) were purchased from the Laboratory Animal Center of Nantong University and used for behavioural studies and primary cultures of mouse Muse cells. Adult male Sprague–Dawley (SD) rats (180 ± 20 g) were used for primary cultures of Muse cells. All animal experiments were approved by the Society for Animal Ethics of Nantong University (permission no. S20220303-005). We used several mouse models of neuropathic pain, including a nerve injury model (CCI and SNI), a STZ-induced diabetes model and a chemotherapy-induced neuropathic pain model. To establish the CCI model, the mice were anaesthetized with isoflurane (RWD), the left sciatic nerve was exposed, and 3 ligatures (7–0 prolene) were placed around the nerve proximal to the trifurcation with 1 millimetre between each ligature. The ligatures were loosely tied until a short flick of the ipsilateral hind limb was observed. The mice in the sham group underwent surgery but not nerve ligation. For SNI surgery, the mice were anaesthetized with isoflurane, and 5.0 silk tight ligation of the tibial and common peroneal nerves was performed, followed by transection and removal of a 3- to 5-mm portion of the nerve. However, the third peripheral branch of the sciatic nerve, the sural nerve, was left intact, and any contact with or stretching of this nerve was carefully avoided. For the diabetes model, the animals were injected intraperitoneally with STZ (Sigma, catalogue: S0130) (80 µg/g in 0.1 M citrate-buffered saline, pH 4.2) for 3 consecutive days. After 2 weeks, we conducted behavioural and blood glucose tests. Mice with blood glucose above 30 mg (16.7 mmol/L) and mechanical pain sensitivity were used for subsequent experiments. For the chemotherapy-induced neuropathic pain model, the animals were injected intraperitoneally with paclitaxel (Selleck, catalogue: S1150) (2 mg/kg in saline) 4 times every other day.

**Behavioural analysis.** The mice were habituated to the testing environment for at least 2 days before baseline testing. The room temperature and humidity remained stable throughout all the experiments. All the behavioural tests were conducted in a blinded manner. To test mechanical sensitivity, we confined the mice to boxes placed on an elevated metal mesh floor and stimulated their hind paws with a series of von Frey hairs with logarithmically increasing stiffness (0.02–2.56 gf; Stoelting), which were presented perpendicularly to the central plantar surface. We determined the 50% paw withdrawal threshold by Dixon's up–down method. Mechanical allodynia after SNI was also assessed according to the frequency of response (expressed as the percentage of response) to a low–threshold von Frey hair (0.16 gf, 10 times). Thermal sensitivity was tested using a Hargreaves radiant heat apparatus (IITC Life Science), with the basal paw withdrawal latency adjusted to 9 – 12 seconds and a cut-off of 20 seconds to prevent tissue damage.

**Drugs and drug administration.** TGF-β–neutralizing Ab, AMD3100 (CXCR4 antagonist), human CCL21/6Ckine and CXCL12/SDF–1α were purchased from R&D Systems; LPS from Invitrogen; TNF-α from Novus; IL-10–neutralizing Ab from BioLegend; and normal rabbit IgG from Cell Signaling Technology. For intrathecal injection, a spinal cord puncture was made with a 29G needle between the L5 and L6 levels to deliver reagents (10 μl) or cells (1 or 2.5 × $10^5$ cells in 10 μl of PBS) to the CSF. Before injection, Muse cells were washed 3 times with 0.01 M PBS, centrifuged for 5 minutes at 1,000 rpm, and then resuspended in PBS. In some cases, Muse cells were incubated with Vybrant CM-DiI Cell-Labeling Solution (Molecular Probes, Life Technologies) for 10 minutes at 37 °C. The cells were then washed 3 times with PBS and resuspended.

**Cell culture.** Muse cell culture was performed according to previous reports [11,14]. In brief, Muse cells were obtained from passage 3 or 4 human, mouse or rat BMSCs. Primary cultures of human BMSCs were isolated from bone marrow that was voluntarily donated by healthy adults. Primary cultures of rodent BMSCs were isolated from adult mice and rats. The BMSCs were subjected to long-term trypsin incubation for 8 h (37 °C, 100% humidity, 5% CO2 in air), followed by vortexing at 2,000 rpm for 3 min (BioCote) and centrifugation at 740 × g for 15 min. To produce M-clusters, individual cells were cultured in MC (MethoCult H4100, StemCell Technologies) culture. For MC culture, culture dishes were purchased from Corning (Ultra Low

Attchment Plates) to avoid the attachment of cells to the bottom of the dish. For suspension culture, the culture medium was α-MEM/20% FBS, and MC was diluted in 20% (vol/ vol) FBS in α-MEM to a final concentration of 0.9%. For adherent culture, the culture medium was α-MEM/20% FBS.

**Flow cytometric analysis.** The properties of the expanded cells were assessed by flow cytometry (BD FACSAria) using PE-conjugated mAbs against CD105 (1:50; eBioscience) and FITC-conjugated mAbs against SSEA-3 (1:50; eBioscience) and analysed using FlowJo software.

**Transwell migration assay.** The migratory ability of the BMSCs and Muse cells was determined using Transwell plates (6.5 mm in diameter with 8μm pore filters; Corning Costar). In brief, $5 \times 10^4$ cells in 100 μl of serum-free medium were added to the upper well, and 600 μl of CXCL12- or CCL21- containing medium and mitomycin C- containing medium were added to the lower well of a Transwell plate. Following incubation for 10 hours (37 °C, 100% humidity, 5% $CO_2$ in air), the number of cells that had migrated to the lower side of the filter was counted under a light microscope(data are presented as the average number of migratory BMSCs and Muse cells in 5 randomly selected fields). Each experiment was performed in triplicate, and the data were averaged for statistical analysis.

**Protein extraction and protein mass apectrometry.** Primary human BMSCs and human Muse cells were washed thoroughly with ice–cold PBS and lysed in buffer containing 50 mM Tris-HCl (pH 7.6), 5 mM EDTA, 50 mM NaCl, 30 mM sodium pyrophosphate, 50 mM NaF, 0.1 mM Na3VO4, 1% (v/v) Triton X-100, 1 mM PMSF, and a protease inhibitor mixture (Roche Applied Science). The lysates were then clarified by centrifugation at 13, 300×g for 30 min at 4 °C, the supernatant was collected, and a BCA assay(Beyotime Biotechnology) was used to determine the protein concentration. Quantitative evaluation of protein expression profiles of two cells was performed using iTRAQ labelling combined with on–line 2D LC/MC/MC proteomics.

**ELISA.** Human TGF-β1 and IL-10 ELISA kits were purchased from R&D Systems. ELISAs were performed using BMSC and Muse cell culture media or cell lysates. Cultured cells were homogenized in lysis buffer containing protease and phosphatase inhibitors. The ELISA was performed according to the manufacturer's instructions.

**Trafficking of Muse cells to DRGs.** To examine the distribution of transplanted CM-Dil–labelled Muse cells following i.t. injection, lumbar spinal cord segments and L4–L6 DRGs were collected. For quantitative analysis of engrafted cells in DRGs, 10 sections (12 μm) from each DRG were examined for the labelled Muse cells.

**IHC.** The animals were deeply anaesthetized with isoflurane and perfused through the ascending aorta with PBS, followed by 4% paraformaldehyde. After perfusion, the spinal cord segments and DRGs were removed and postfixed in the same fixative overnight. Spinal cord and DRG sections were cut with a cryostat and processed for immunofluorescence. The sections were incubated overnight at 4 °C with the following primary Abs: mouse GFAP (1:1,000; EMD Millipore) and rabbit IBA-1 (1:1,000; Wako Chemicals USA Inc.). In some cases, Hoechst (Sigma) was used to stain the cell nuclei. The stained sections were examined under a Nikon fluorescence microscope or a Zeiss confocal microscope. The intensity of fluorescence was analysed using Photoshop CS6 (PS).

**qPCR.** Total RNA was isolated from spinal dorsal horn tissues and DRGs of the L4–L5 segments using the TRIzol RNA isolation system (Sigma). cDNA was synthesized from total RNA using a PrimescriptTM RT Master Mix kit (TAKARA) following the supplier's instructions. We performed gene-specific mRNA analyses using the SYBR Premix Ex TaqTM Real–Time PCR system (TAKARA). The primer sequences are described in Table s1. The primer efficiency was obtained from the standard curve and integrated for the calculation of relative gene expression, which was based on the real–time PCR threshold values of different transcripts and groups.

**Statistical analyses.** All mice were randomly assigned to different groups. The data were analyzed by two independent researchers in a double-blind manner. All data from mice were included except for the screening of mice with successful STZ model construction. For immunohistochemistry or behavioral analysis, we conducted two or three independent replications. The experimental data was analysed using GraphPad

Prism 8.0 software. The measured data are expressed as the mean ± standard error of the mean (SEM). Student's t-test (two groups) or ANOVA (one-way or two-way) were used to compare the differences between groups, followed by the Bonferroni correction. A value of P < 0.05 was considered statistically significant.

**Results**

**Characterization of Muse cells in human and rodents**

As previously described[11], Muse cells were obtained from cultured BMSCs by long–term trypsin incubation. Eight hours of trypsin treatment resulted in many dead cells, and the surviving cells were collected and cultured under suspension culture conditions[11]. These cells generated cell clusters after 1 week and could be subcultured many times (Figure 1A). These cells were confirmed to be Muse cells by their specific cell surface markers (SSEA-3 and CD105; Figure 1B, Figure s1A). Muse cells can also be cultured in adherent media and have a cell shape similar to that of BMSCs (Figure 1C). In this study, we cultured human, rat and mouse Muse cells and found that they had the same characteristic features under suspension culture and adherent culture conditions (Figure 1A-C).

**iTRAQ analysis of Muse cells and BMSCs**

Proteomic strategies represent powerful tools for the global investigation of many cellular proteins. To explore the potential functions of Muse cells, quantitative proteomic analysis based on isobaric tags for relative and absolute quantitation (iTRAQ) labelling was conducted to obtain an unbiased view of the proteomic profiles of hMuse cells and hBMSCs. With a false discovery rate (FDR) of less than 1%, a total of 4368 proteins were identified. Among the 4,368 identified proteins, 3,624 proteins were identified by two or more unique peptides, and the remaining 744 proteins were identified by one unique peptide. To characterize the differences between Muse cells and BMSCs, altered proteins were determined according to their relative protein expression (Muse cells/BMSCs). According to the threshold values for downregulated and upregulated proteins ($\leq 0.20$ and $\geq 4.44$, respectively and 90% confidence interval), a total of 105 downregulated and 59 upregulated proteins were found in Muse cells (Figure 1D). Thus, the protein maps of Muse cells and BMSCs were 96.25% similar. The differentially expressed proteins were classified into 14 categories according to their main biological functions collected from relevant literature in PubMed, including metabolism, cell proliferation and growth, transcription and translation, transport, migration, cytoskeleton, differentiation, cell adhesion, apoptosis, immune response, signal transduction, development, and the cell cycle (Figure 1E).

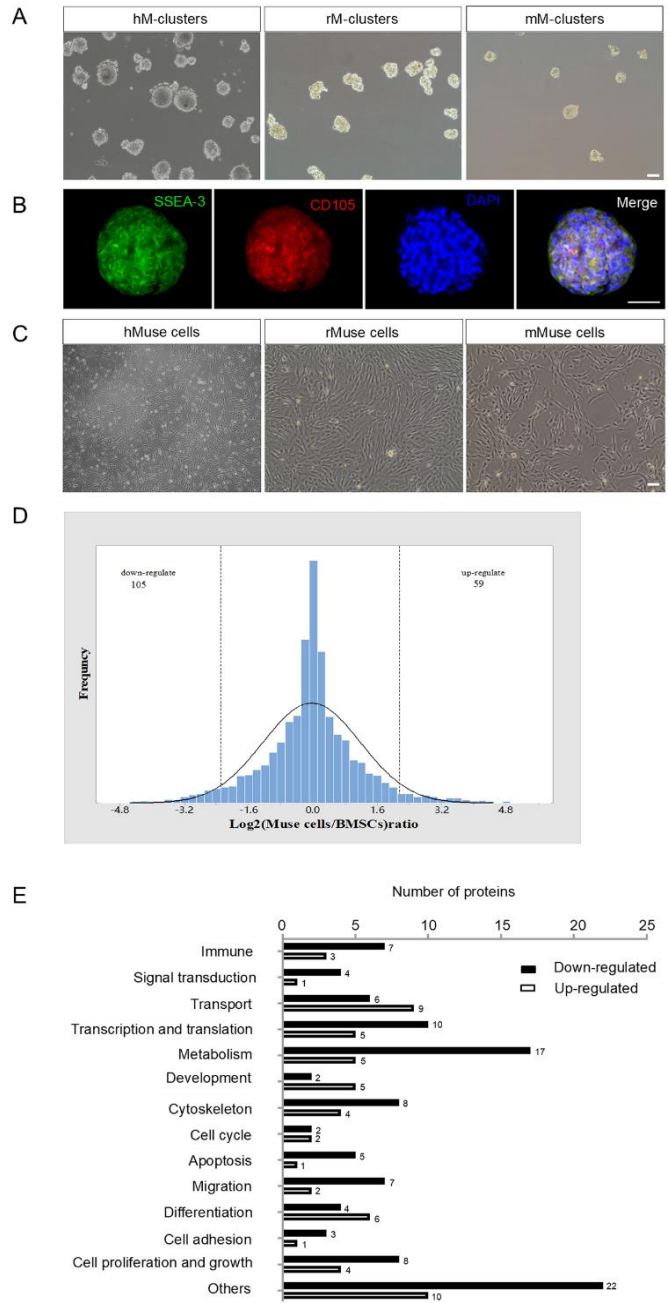

**Figure 1: Characterization of Muse cells.**

(A) Representative images of single Muse cell from human, rat and mouse formed cell cluster after 1 week of suspension culture. Scale bar, 100 μm. (B) Human Muse cell cluster expressed SSEA-3 and CD105. Scale bar, 100 μm. (C) Representative images of Muse cell from human, rat and mouse in adherent culture for 1 week. Scale bar, 100 μm. (D) Distribution of mean ratios of 4368 identified proteins, as measured by three independent iTRAQ experiments. Ratios were calculated as human Muse cells versus human BMSCs. The 90% confidence intervals are indicated by vertical lines in the plot. A total of 105 down-regulated and 59 up-regulated proteins were found in Muse cells. (E) Functional categories of differentially expressed proteins between human Muse cells versus human BMSCs.

**Inhibition of neuropathic pain in mice by a single intrathecal injection of Muse cells**

To test the hypothesis that Muse cells alleviate neuropathic pain, various mouse models including nerve injury, diabetes and chemotherapy-induced neuropathic pain were used. First, the analgesic efficacy of Muse cells from different species was examined in a sciatic nerve chronic constriction injury (CCI)-induced persistent neuropathic pain mouse model [15,16]. Twelve days after CCI, the same number ($2.5 \times 10^5$ cells) of human, rat or mouse Muse cells were intrathecally injected, all of which rapidly (<1 day) and long-lastingly (>42 days) inhibited CCI-induced mechanical allodynia (Figure 2A) and thermal hyperalgesia (Figure 2B). There was no significant difference in the analgesic effects among the three types of Muse cells. Notably, these experiments were performed in wild-type mice, and no immunosuppressive agents were used.

Diabetes associated with peripheral neuropathy is a serious health problem. The streptozotocin (STZ)-induced diabetes model elicits mechanical allodynia and heat hypoalgesia and is recognized as one of the most difficult types of sensorial neuropathy to treat[17]. Behavioural signs of mechanical allodynia and thermal hypoalgesia were evident 2 weeks after diabetes model induction. Mechanical allodynia was maintained for 8 weeks, at which time diabetic mice showed a gradual loss of mechanical sensitivity, while thermal hypoalgesia persisted during the experimental period of 15 weeks. To determine whether Muse cells induce therapeutic effects on diabetic sensory neuropathy, mice were treated with human Muse cells ($2.5 \times 10^5$) (treated with PBS as vehicle group) via the i.t. route 20 days after diabetes induction, when sensory neuropathy was fully established. Two weeks after administration, neuropathic mice treated with Muse cells exhibited antinociceptive effects in response to mechanical stimuli (Figure 2C). The antinociceptive effect of Muse cells was progressive, peaking three weeks after treatment, when complete reversal of mechanical allodynia was achieved. Importantly, the progression of sensory neuropathy, indicated by the late loss of mechanical sensitivity, was completely prevented in Muse cell-treated mice. Additionally, Muse cell treatment reversed the thermal hypoalgesia of neuropathic model mice from 10 days after administration until 8 weeks after treatment (Figure 2D).

Paclitaxel-induced peripheral neuropathy is a neuropathic pain model that represents a frequent and serious consequence of chemotherapy agents[18]. In this model, paclitaxel–induced mechanical allodynia in mice reached the maximum level 1 day after the induction of chemotherapy, lasted for approximately 3 weeks, and then gradually returned to normal levels 8 weeks after induction (Figure 2E). To determine whether Muse cells induce therapeutic effects in a chemotherapy-induced neuropathic pain model, mice were treated with human Muse cells ($2.5 \times 10^5$) via the i.t. route 1 day after paclitaxel (2 mg/kg, 4 times every other day) treatment. The antinociceptive effect of Muse cells on mechanical stimuli was observed 5 h after administration and gradually peaked at 5 weeks before being maintained until the end of the evaluation period (Figure 2E; $p < 0.05$).

Taken together, these results indicate that the intrathecal injection of Muse cells has long-term analgesic effects on neuropathic pain model mice.

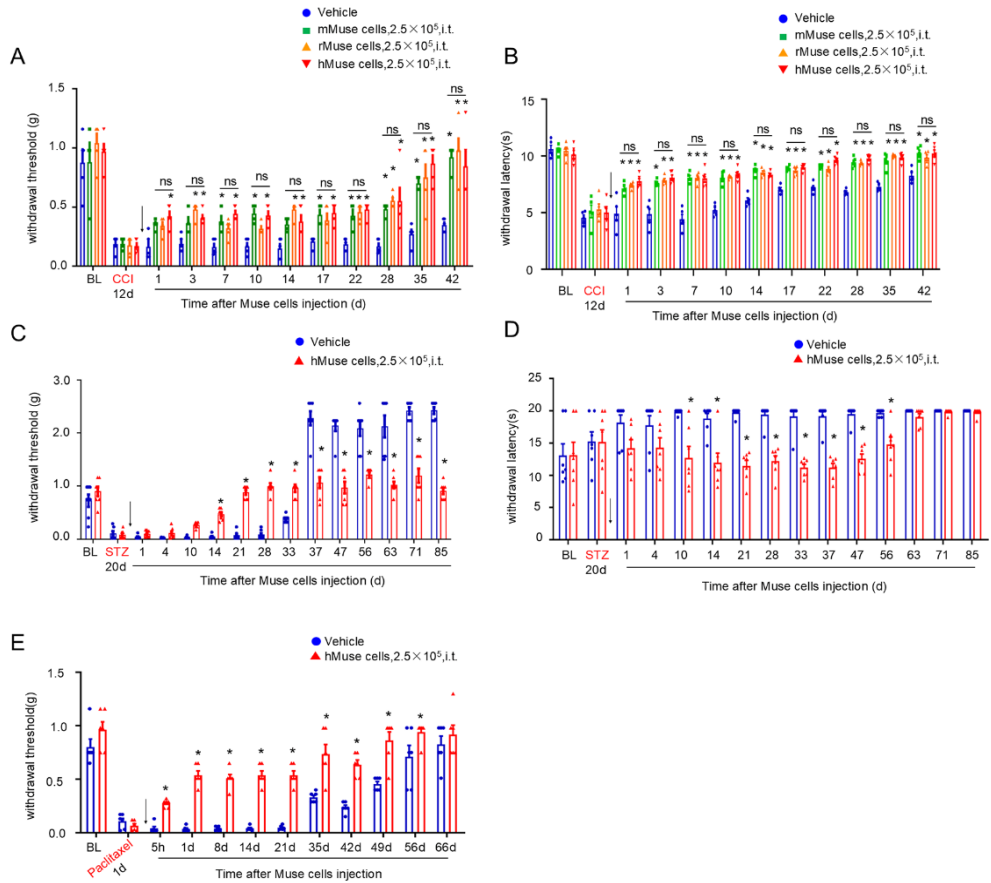

**Figure 2: A single intrathecal injection of Muse cells inhibit neuropathic pain in mice.**

(A, B) Intrathecal injection of the same amount ($2.5 \times 10^5$ cells) of human, rat or mouse Muse cells, 12 days after CCI, all produced a rapid (<1 day) and long-lasting (>42 days) inhibition of CCI-induced mechanical allodynia (A) and thermal hyperalgesia (B). n = 4 or 5 mice/group. (C, D) Intrathecal injection of human Muse cells inhibited STZ-induced mechanical allodynia (C) and thermal hypoalgesia (D). n = 7 mice/group. (E) Intrathecal injection of human Muse cells inhibited paclitaxel-induced mechanical allodynia. n = 6 mice/group. *$P < 0.05$, compared with vehicle; statistical significance was determined by Two way ANOVA, followed by Bonferroni's post-hoc test. All data are expressed as the mean ± SEM. BL, baseline.

**Muse cells have a stronger and more stable analgesic effect than BMSC cells**

To compare the analgesic effects of BMSCs and Muse cells on neuropathic pain in mice, three doses ($1\times10^4$, $5\times10^4$ and $2.5\times10^5$) of primary cultured human BMSCs or human Muse cells were intrathecally injected into the mice 5 days after spared nerve injury (SNI). As shown in Figure 3A, intrathecal injections of high-dose ($2.5\times10^5$) BMSCs or Muse cells produced a similar rapid and long-lasting analgesic effects, and there was no significant difference between the two treatments. However, the analgesic effects of injections of low-dose ($1\times10^4$) and middle-dose ($5\times10^4$) Muse cells were stronger and more durable than those of the same dose of BMSCs.

Next, we compared the analgesic effects of cultured human BMSCs and Muse cells at different passages. Passage 0 (P0) and passage 20 (P20) BMSCs and the same passage number of Muse cells (Additional file 1: Figure S1B) were intrathecally injected into SNI model mice on Day 5. At the same time, the PBS treatment group was used as the vehicle group. Similar to the above results, both P0 BMSCs and P0 Muse cells effectively reversed SNI-induced mechanical allodynia for a long period of time (Figure 3B). Interestingly, the intrathecal treatment of P20 Muse cells into SNI mice also produced effective and long-term analgesic effects, whereas P20 BMSCs produced only mild and transient analgesic effects (Figure 3B). Notably, there was no significant difference in the analgesic effects of P0 Muse cells and P20 Muse cells, which indicates that the analgesic ability of Muse cells is very stable. In summary, these results suggest that Muse cells have stronger and more stable analgesic effects than BMSCs do, especially after long-term culture and passaging.

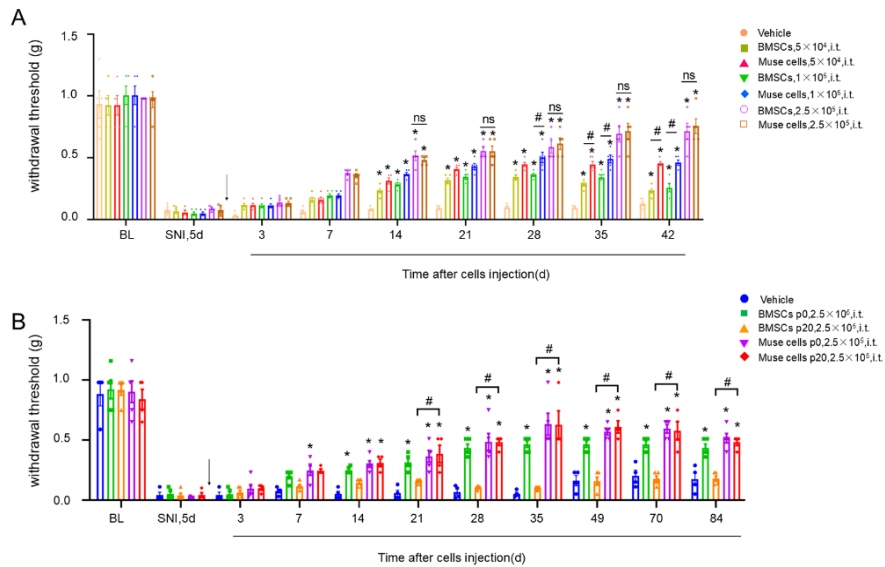

**Figure 3: Muse cells have a stronger and more stable analgesic effect than BMSC cells.**

(A) Intrathecal injection of different number of human BMSCs or Muse cells all inhibited SNI-induced mechanical allodynia. The analgesic effects of injection of low- ($1\times10^4$) and middle-does ($5\times10^4$) Muse cells were stronger and more durable than the same dose of BMSCs. n = 5 mice/group. (B) The analgesic effects of different passages (P0 and P20) of human BMSCs and Muse cells on SNI-induced mechanical allodynia. P20 Muse cells produced effective and long-term analgesic effects, whereas P20 BMSCs only produced mild and transient analgesic effects. n = 4 or 5 mice/group. *$P < 0.05$, compared with vehicle; #$P < 0.05$; statistical significance was determined by Two way ANOVA, followed by Bonferroni's post-hoc test. All data are expressed as the mean ± SEM. BL, baseline.

**Intrathecal Muse cells inhibited SNI–induced DRG neuron damage and spinal glial activation**

SNI-induced axonal injury elicits DRG neuron injury and spinal glial activation, which contribute to the induction and persistence of neuropathic pain[16,19]. First, we investigated whether the intrathecal injection of Muse cells can protect injured DRG neurons in SNI mice. Activating transcription factor 3 (ATF3) is elicited in various tissues in response to stress and can be used as an indicator to assess the degree of damage. Compared with the positive rate of less than 1% in the sham–operated mice, SNI-induced ATF3 expression in L4-L5 DRG neurons 10 days after nerve injury significantly increased to 45% (Figure 4A), and a single intrathecal injection of Muse cells 5 days after SNI reduced the percentage of ATF3-positive neurons to 19% at the same timepoint (Figure 4A).

Next, we investigated the inhibitory effects of intrathecal Muse cells on spinal glial cell activation[20]. Immunoreactivity of both GFAP (an astrocyte marker) and IBA-1 (a microglial marker) were markedly increased in the dorsal horn of spinal cord 10 days after SNI (Figure 4B, C) and were strongly suppressed at the same timepoint after Muse cell injection at 5 days after SNI (Figure 4B, C). Notably, the morphological changes in spinal astrocytes and microglia induced by SNI were also inhibited by Muse cell treatment. Since activated spinal glial cells participate in the regulation of neuropathic pain through the production of several inflammatory cytokines[21,22], such as IL-1β, IL-6 and TNF-α, we further examined the mRNA levels of these inflammatory cytokines in the spinal cord dorsal horns of sham surgery and SNI model mice with or without Muse cell treatment. The qPCR results revealed that intrathecal Muse cells inhibited the increase in IL-1β, IL-6 and TNF-α levels induced by SNI in the dorsal horn of spinal cord (Figure 4D). Collectively, these data indicate that the intrathecal injection of Muse cells reduces SNI-induced DRG neuron damage and spinal glial activation.

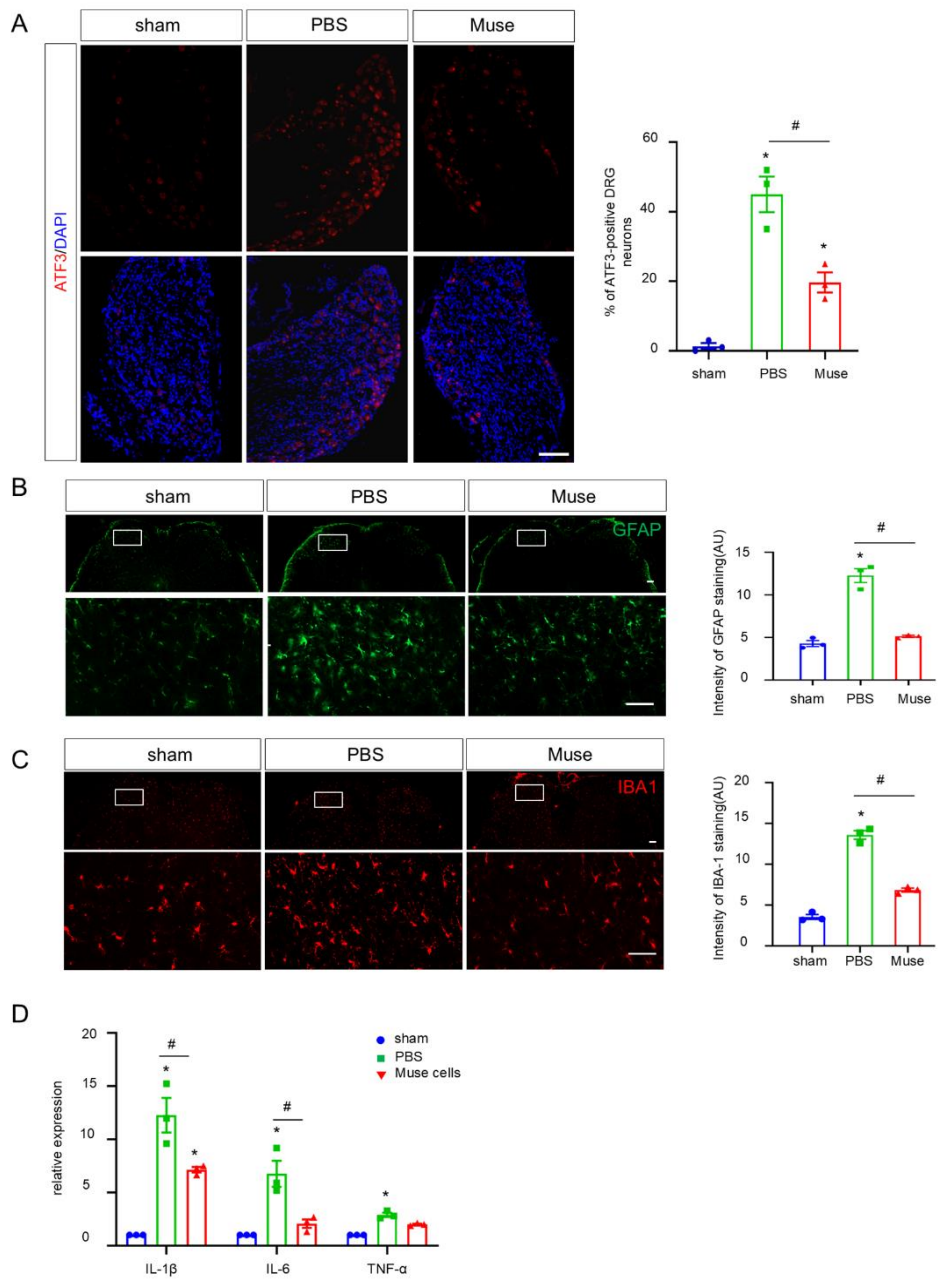

**Figure 4: Intrathecal Muse cells inhibited SNI-induced DRG neurons damage and spinal glial activation.**

(A) Left: Double staining of the DRG neuron damage marker ATF3 (red) and the nucleus marker DAPI (blue) in ipsilateral L5 DRG 10 days after SNI. Right: Quantification results of ATF3 staining in DRGs. Compared the PBS group, intrathecal injection of Muse cells ($2.5 \times 10^5$ cells), given 5 days after SNI, reduced the ATF3 expression. Scale bars: 50 μm. *$P < 0.05$, compared with the sham, #$P < 0.05$; statistical significance was determined One way ANOVA, followed by Bonferroni's post-hoc test. n = 3 mice/group. All data are expressed as the mean ± SEM. (B, C) Intrathecal Muse cells ($2.5 \times 10^5$ cells), given 5 days after SNI, reduced SNI-induced GFAP (B, the astrocyte marker) and IBA-1 (C, the microglial marker) expression in the L4–L5 spinal cord dorsal horn. Bottom panels are enlarged images of the top panels. Scale bar, 50 μm. Graph in B and C shows the quantification of GFAP and IBA-1 staining. *$P < 0.05$, compared with the sham group; #$P < 0.05$; statistical significance was determined by One way ANOVA, followed by Bonferroni's post-hoc test. n = 3 mice/group. (D) qPCR showing the expression levels of IL-1β, IL-6, and TNF-α mRNAs in the spinal cord dorsal horn of sham and SNI mice with or without Muse cells treatment. *$P < 0.05$, compared with the sham group; #$P < 0.05$; n = 3 mice/group. Statistical significance was determined by One way ANOVA, followed by Bonferroni's post-hoc test. All data are expressed as the mean ± SEM.

**Muse cells inhibit neuropathic pain via TGF-β and IL-10 secretion**

To explore the molecular mechanism of analgesia in Muse cells, we tested the secretion of anti–inflammatory factors by Muse cells. The cytokines TGF-β and IL-10 have been widely demonstrated to have anti-inflammatory effects and participate in the beneficial effects of BMSC treatment[3,4]. Our previous report revealed that BMSCs release TGF-β but not IL-10 to inhibit neuropathic pain. In this study, we compared the expression and secretion of TGF-β and IL-10 in Muse cells and BMSCs. The TGF-β ELISA results showed lower levels of TGF-β expression but similar levels of TGF-β secretion in Muse cells compared with BMSCs (Figure 5A). Importantly, the IL-10 ELISA results revealed that, compared with that in BMSCs, the expression level of IL-10 in Muse cells was 3-fold greater, and the secretion level was 40-fold greater (Figure 5A). Next, we examined the release of TGF-β and IL-10 in cerebrospinal fluid (CSF) collected from sham surgery and SNI mice with or without Muse cell treatment. TGF-β and IL-10 release in the CSF did not increase in SNI mice but significantly increased 4 days after they received an intrathecal administration of Muse cells (Figure 5B).

Behavioural tests further revealed that TGF-β and IL-10 are involved in the analgesic effect of Muse cells. As shown in Figure 5C, the intrathecal injection of TGF-β- and IL-10- neutralizing antibodies partially reversed the analgesic effect of Muse cells, and when these two antibodies were injected together, the effect of reversing the analgesic effect was more prominent.

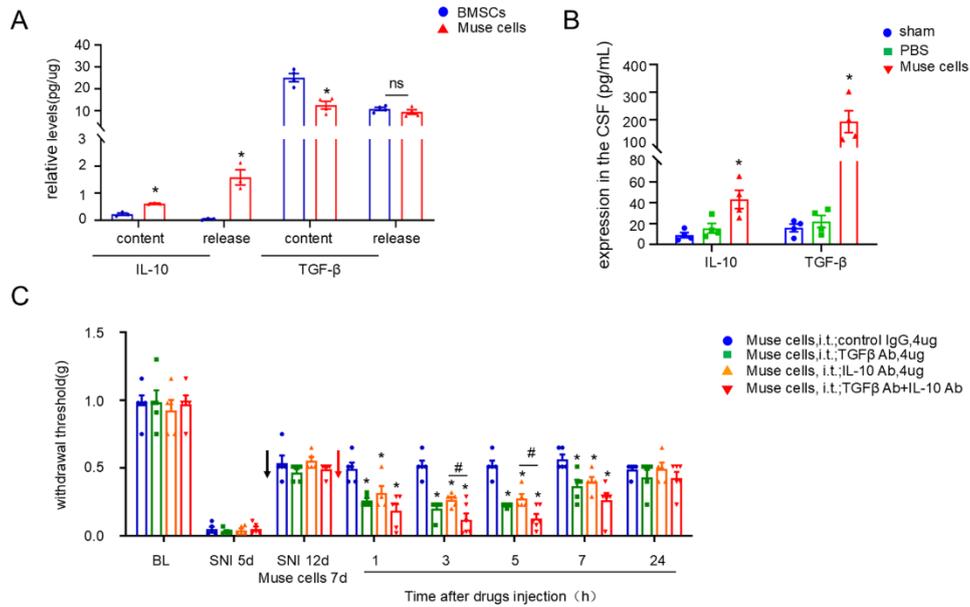

**Figure 5: Muse cells release IL-10 and TGF-β to inhibit neuropathic pain in SNI mice.**

(A) ELISA analysis showing IL-10 and TGF-β release in BMSC and Muse cells culture medium. Compared with BMSCs, a lower expression but similar secretion levels of TGF-β were observed in Muse cells. However, the expression level of IL-10 in Muse cells was 3-fold higher and the secretion level was 40-fold higher than BMSCs. *P < 0.05, compared with BMSCs; statistical significance was determined by Student's t test, n = 3 or 4 separate cultures. (B) ELISA analysis showing increased IL-10 and TGF-β release in CSF 8 days after SNI and 4 days after i.t. delivery of $2.5 \times 10^5$ Muse cells. *P < 0.05, compared with sham and PBS i.t.; statistical significance was determined by One way ANOVA, followed by Bonferroni's post-hoc test. n = 4 mice/group. (C) Reversal of Muse cells-induced inhibition of mechanical allodynia by IL-10 andTGF-β–neutralizing Abs (4 μg, i.t.), but not by control IgG (4 μg, i.t.). *P < 0.05, compared with the control IgG group; #P < 0.05; statistical significance was determined by Two way ANOVA, followed by Bonferroni's post-hoc test. n = 5 or 6 mice/group.

**CCL21/CCR7 axis controls Muse cell migration**

Finally, we investigated the distribution of Muse cells after their intrathecal injection into CCI mice. In accordance with our previous methods, the fluorescent dye CM-Dil was used to label the injected Muse cells. Four weeks after injection, we examined the distribution of Muse cells on both sides of the L5 DRG, which is the most severe segment of DRG damage caused by the CCI model. Similar to the previous results of BMSC injection, very few Muse cells were detected in the L5 DRG on the normal control side (data not shown), and many Muse cells were detected on the injured side (Figure 6A). The number of Dil-labelled Muse cells peaked on Day 3, followed by a gradual decrease. Importantly, Dil-labelled Muse cells were still detectable in the DRGs up to Day 84 post-implantation, supporting their ability to survive in the DRG for an extended period of time(Figure 6A, 6B).

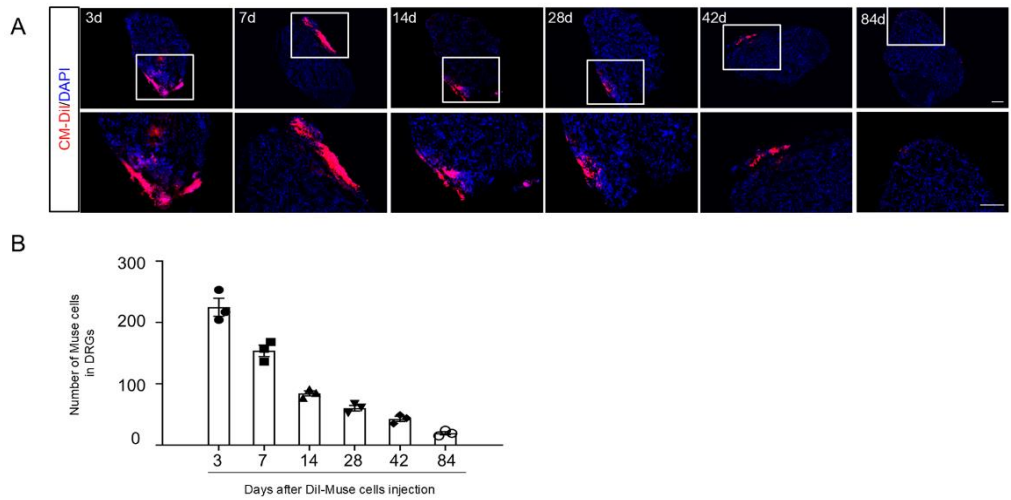

**Figure 6: Long-term survival of CM-Dil–labeled Muse cells in ipsilateral L5 DRGs following i.t. injection in CCI mice.**

(A) Localization of CM-Dil–labeled Muse cells in ipsilateral L5 DRGs 3–84 days after i.t. injection. Scale bars, 100 μm. Bottom panels are enlarged images of the top panels. Note that Muse cells were mainly localized at DRG borders.

(B) Number of CM-Dil–labeled Muse cells in L5 DRGs 3–84 days after i.t. Muse cell injection, given 4 days after CCI. n = 3 mice/group.

To study the mechanism of Muse cell chemotaxis in CCI mice, we first measured the expression of all known chemokine receptors, including CXCR1-6, CCR1-10, XCR1 and CX3CR1, in Muse cells. Among all the examined receptors, the expression of CCR7 was unexpectedly much greater than that of the other receptors (Figure 7A). Since the ligands of CCR7 are CCL19 and CCL21[23-25], we next detected the expression of these two ligands in DRG neurons. The RT-PCR results revealed that in the DRGs of mice 5 days after CCI, the expression of CCL21 in the ipsilateral DRG was increased 1,000-fold compared with that in the sham group. In contrast, very low expression of CCL19 mRNA was detected in DRG tissues (Figure 7B). To further examine the role of the CCL21/CCR7 axis in regulating the migration of Muse cells, a Transwell migration assay was used, and BMSCs were used as a control. Interestingly, the number of migrating Muse cells was three times greater than that of BMSCs under normal culture conditions (Figure 7C, D). CXCL12 induced the migration of only BMSCs but not Muse cells, while CCL21 induced both Muse cell and BMSC migration (Figure 7C, D). These results indicate that the CCL21/CCR7 axis controls Muse cell migration.

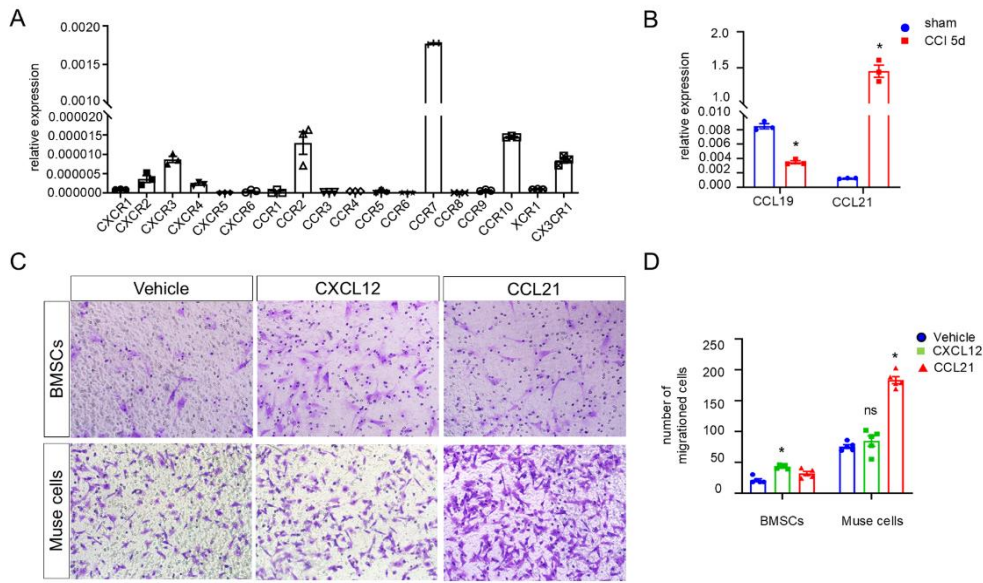

**Figure 7: CCL21/CCR7 axis controls Muse cells migration.**

(A) qPCR showing the expression levels of all known chemokine receptors in Muse cells, including CXCR1-6, CCR1-10, XCR1 and CX3CR1. Note, the expression of CCR7 was unexpectedly much higher than other receptors. n = 3 cultures. (B) qPCR showing CCL19 and CCL21 mRNA expression in ipsilateral L5 DRGs on days 5 after CCI. The expression of CCL21 in CCI mice was increased more than 1000-fold than the sham group. *$P < 0.05$; n = 3 mice/group. (C) Chemotaxis (Transwell invasion) assay showing the effects of CXCL12 and CCL21 on the migration of BMSCs and Muse cells. Note, CXCL12 only induced migration of BMSCs but not Muse cells, while CCL21 induced both Muse cells and BMSCs migration. *$P < 0.05$, compared with the vehicle group (no treatment); statistical significance was determined by One way ANOVA, followed by Bonferroni's post-hoc test. n = 5 wells from separate cultures.

**Discussion**

Neuropathic pain is a serious and growing health problem. Currently, there are no drugs or treatments that can completely and definitively alleviate neuropathic pain [26]. Accumulating evidence indicates that the transplantation of BMSCs can relieve neuropathic pain[16]. However, BMSCs are heterogeneous and contain many different cell subtypes, making it unclear which cell subsets play an analgesic role[27]. Here, we identified Muse cells, a subpopulation of BMSCs with potential value in treating neuropathic pain in various mouse models, including nerve injury, diabetes, and chemotherapy–induced neuropathic pain. We also provided evidence that this analgesic effect is produced by the secretion of TGF-β and IL-10 from Muse cells. Additionally, the analgesic effect of Muse cells did not change after 20 passages.

The most valuable aspect of this study is the discovery of a subpopulation of BMSCs that can relieve pain for a long period of time after intrathecal injection. BMSCs are mixed cells containing various subpopulations, and previous studies have rarely focused on a single subpopulation. Muse cells are a newly discovered cell subgroup, and many functions of Muse cells are still unclear[28]. In this study, mass spectrometry analysis revealed that the protein compositions of Muse cells and BMSCs were more than 95% similar, suggesting that Muse cells may have physiological functions similar to those of BMSCs. Previous studies of Muse cells have focused mainly on their stem cell characteristics, in which Muse cells were found to be able to repair damaged tissues by differentiating into specific types of cells[12,13], and there are few studies describing their anti-inflammatory ability. Here, we compared the anti-inflammatory and analgesic abilities of Muse cells and BMSCs. Intrathecal injection of both types of cells reduced nerve injury-induced neuropathic pain for a long time, and both reduced the inflammatory response and glial cell activation, protecting damaged neurons. Additionally, Muse cells had three advantages over BMSCs in terms of analgesic effects. First, Muse cells had stronger analgesic effects than did low-dose BMSCs ($5\times10^4$ cells, i.t.), effectively reducing the number of cells needed for clinical treatment. Second, Muse cells simultaneously secreted TGF-β and IL-10 for analgesia, which has more applications in clinical therapy[29-31]. Third, the analgesic ability of Muse cells remained stable after long-term subculture, whereas the analgesic ability of BMSCs appeared to be significantly weakened. This advantage of Muse cells will provide a key benefit and convenience in the development of commercial cellular drugs with stable analgesic effects.

The main purpose of this study was to facilitate the application of BMSCs in the treatment of neuropathic pain. Currently, because of the safety of cell culture, most clinical treatments for BMSCs are simple separation by flow cytometry after bone marrow extraction, followed by autologous BMSC transplantation[32]. Patients need to endure considerable pain to extract a sufficient amount of bone marrow to obtain the number of BMSCs required for transplantation[33]. Compared with patients with other serious and dysfunctional diseases, such as nerve injury, liver disease and arthropathy, patients with chronic pain are less willing to receive autologous BMSC transplantation.

BMSCs are a group of cells consisting of various subtypes[34], and no unified criterion has been established to identify BMSCs and no specific marker molecule for BMSCs has been identified to date[35]. Because BMSCs obtained under different culture conditions are different, the data from different laboratories are not comparable. In addition, even in the same laboratory, BMSCs may undergo significant changes in cell characteristics[36], such as loss of multilineage differentiation and analgesia, after multiple passages [37]. Therefore, these characteristics of BMSCs make it difficult to establish strict quality control standards for commercially available BMSCs. In contrast, Muse cells constitute a homogeneous cell group with consistent characteristics[11]. In this study, our results confirmed that single Muse cells rapidly proliferated into pellets in suspension culture and that Muse cells still had stable and strong analgesic effects after multiple passages. Combined with other characteristics of Muse cells, such as their nontumorigenic and strong immunosuppressive properties, Muse cells are good candidates for commercial analgesic cell therapy.

Previous studies have reported that Muse cells transplanted via the systemic route can effectively migrate to damaged tissues and can spontaneously differentiate into cells compatible with homing tissues[38]. In this study, we demonstrated a paracrine mechanism by which intrathecal Muse cells target CCL21-producing DRGs to achieve long-term relief of neuropathic pain via the secretion of TGF-β and IL-10. A small intrathecal injection space allows a small amount of cytokines secreted by Muse cells to reach therapeutic concentrations. Some clinical trials have shown that intrathecal

BMSCs are safe and do not cause health problems for at least 12 months[39]. The intrathecal route not only requires far fewer cells than the intravenous route but also avoids systemic immune responses to transplanted cells because of the blood–spinal/brain barrier. Our previous results revealed that intrathecally injected BMSCs act directly on the spinal cord and DRG and can survive in these tissues for a long time, providing sustained analgesic and neuroprotective effects. In comparison, transplanted BMSCs via the systemic route are mostly confined to the lung and survive for only a few days[40].

Overall, Muse cells can be used as a stable, standardized commercial stem cell source for allogeneic transplantation for the treatment of neuropathic pain. The intrathecal injection of Muse cells may provide an efficient, long–term, safe and inexpensive way to treat chronic pain in future clinical trials.

**Conclusion**

Muse cells can reverse different types of neuropathic pain and maintain pain relief for a long period of time. Intrathecal Muse cells migrate to injured DRGs through the CCL21–CCR7 chemotactic axis and exert analgesic effects through TGF-β and IL-10, which may provide a new research direction for the clinical treatment of neuropathic pain.

**Abbreviations**

*Muse:* Multilineage differentiation stress enduring

*BMSC:* Bone marrow stromal cells

*CCI:* Chronic constriction injury model

*SNI:* Spared nerve injury model

*STZ:* Streptozotocin

*TGF-β:* Transforming growth factor-beta

*CCL21:* C–C Motif Chemokine 21

*CXCL12:* C–X–C motif chemokine ligand 12

*TNF-α:* Tumor necrosis factor-alpha

*IL-10:* Interleukin (IL)-10

*CSF:* Cerebral spinal fluid

*CD105:* Endoglin, cell membrane glycoprotein

*SSEA-3:* Stage–specific embryonic antigen–3

*DRG:* Dorsal root ganglion

*ATF3:* Activating Transcription Factor 3

*GFAP:* Glial fibrillary acidic protein

*IBA-1:* Ionized calcium binding adapter molecule 1

*IL-1β:* Interleukin (IL)-1-beta

*IL-6:* Interleukin (IL)-6

*CCR7:* C–C chemokine receptor type 7

# Supplementary Information

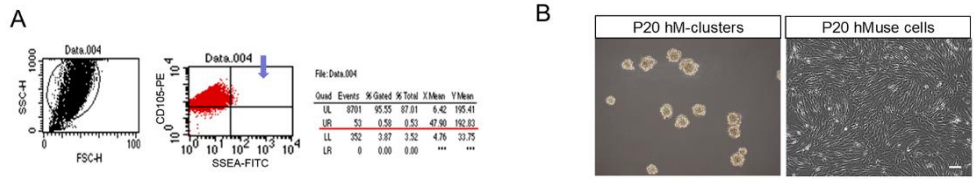

**Supplementary Material 1 : Figure s1: Characterization of Muse cells.**

(A) According to the characteristics of Muse cells, flow cytometry sorting was used to analyze the SSEA-3+/CD105+ cells, and about 0.58% of them were Muse cells.(B) Representative images of P20 hMuse cell after 1 week of suspension culture (left) and adherent culture (right). Scales, 100 μm.

**Supplementary Material 1 : Table s1**

Table s1 Primers used for quantitative reverse transcription-polymerase chain

| Gene | Sequence | Product length (bp) |
|---|---|---|
| mIL-1β | Forward: 5′-TGTCTTGGCCGAGGACTAAG-3'<br>Reverse: 5′-TGGGCTGGACTGTTTCTAATG-3' | 110 |
| mIL-6 | Forward: 5′-TCCATCCAGTTGCCTTCTTGG-3'<br>Reverse: 5′-CCACGATTTCCCAGAGAACATG-3' | 165 |
| mTNF-α | Forward: 5′-CCCCAAAGGGATGAGAAGTT-3'<br>Reverse: 5′-CACTTGGTGGTTTGCTACGA-3' | 132 |
| mCCL19 | Forward: 5′-TGCCTGCTGTTGTGTTCACC-3'<br>Reverse: 5′-GGCTCCTTCTGGTGCTGTTG-3' | 135 |
| mCCL21 | Forward: 5′-GGTCCTGGCTCTCTGCATTC-3'<br>Reverse: 5′-GCTTCCTGTAGCCTCGGACA-3' | 116 |
| mGAPDH | Forward: 5′-TCCATGACAACTTTGGCATTG-3'<br>Reverse: 5′-CAGTCTTCTGGGTGGCAGTGA-3' | 72 |
| hCXCR1 | Forward: 5′-GGCTGCTGGGGACTGTCTATGAAT-3'<br>Reverse: 5′-GCCCGGCCGATGTTGTTG-3' | 383 |
| hCXCR2 | Forward: 5′-CCGCCCCATGTGAACCAGAA-3'<br>Reverse: 5′-AGGGCCAGGAGCAAGGACAGAC-3' | 427 |
| hCXCR3 | Forward: 5′-CCACCCACTGCCAATACAAC-3' | 379 |

| | Reverse: 5′-CGGAACTTGACCCCTACAAA-3' | |
|---|---|---|
| hCXCR4 | Forward: 5′-ACTACACCGAGGAAATGGGCT-3'<br>Reverse: 5′-CCCACAATGCCAGTTAAGAAGA-3' | 133 |
| hCXCR5 | Forward: 5′-GGTCACCCTACCACATCGTC-3'<br>Reverse: 5′-GCCATTCAGCTTGCAGGTATTG-3' | 80 |
| hCXCR6 | Forward: 5′-GACTATGGGTTCAGCAGTTTCA-3'<br>Reverse: 5′-GGCTCTGCAACTTATGGTAGAAG-3' | 169 |
| hCCR1 | Forward: 5′-ACCTGCAGCCTTCACTTTCCTCA-3'<br>Reverse: 5′-GGCGATCACCTCCGTCACTTG-3' | 327 |
| hCCR2 | Forward: 5′-TACGGTGCTCCCTGTCATAAA-3'<br>Reverse: 5′-TAAGATGAGGACGACCAGCAT-3' | 120 |
| hCCR3 | Forward: 5′-ATGCTGGTGACAGAGGTGAT-3'<br>Reverse: 5′-AGGTGAGTGTGGAAGGCTTA-3' | 298 |
| hCCR4 | Forward: 5′-GAAGAAGAACAAGGCGGTGAAGAT-3'<br>Reverse: 5′-ATGGTGGACTGCGTGTAAGATGAG-3' | 391 |
| hCCR5 | Forward: 5′-CATCTGCTACTCGGGAATCCT-3'<br>Reverse: 5′-AGCCCACTTGAGTCCGTGTC-3' | 445 |
| hCCR6 | Forward: 5′-CTCCAGGCTATTTGTACCGATTG-3'<br>Reverse: 5′-CACTGCCCAGAATGGGAGAG-3' | 181 |
| hCCR7 | Forward: 5′-GTGCCCGCGTCCTTCTCATCAG-3'<br>Reverse: 5′-GGCCAGGACCACCCCATTGTAG-3' | 353 |

| Gene | Primers | Size (bp) |
|---|---|---|
| hCCR8 | Forward: 5′-GCCGTGTATGCCCTAAAGGT-3'<br>Reverse: 5′-ATGGCCTTGGTCTTGTTGTG-3' | 299 |
| hCCR9 | Forward: 5′-CACTGTCCTGACCGTCTTTGTCT-3'<br>Reverse: 5′-CTTCAAGCTTCCCTCTCTCCTTG-3' | 292 |
| hCCR10 | Forward: 5′-TGCTGGATACTGCCGATCTACTG-3'<br>Reverse: 5′-TCTAGATTCGCAGCCCTAGTTGTC-3' | 302 |
| hXCR1 | Forward: 5′-TGACCATCCACCGCTACC-3'<br>Reverse: 5′-ATCTGGGTCCGAAACAGC-3' | 397 |
| hCX3CR1 | Forward: 5′-TCCTTCTGGTGGTCATCGTG-3'<br>Reverse: 5′-CCAGCCTCAGATCCTTCCTC-3' | 123 |
| hCXCL12 | Forward: 5′-CGCTCTGCATCAGTGACGGTAAG-3'<br>Reverse: 5′-CGTTGGCTCTGGCGACATGG-3' | 87 |
| hGAPDH | Forward: 5′-GGGGAGCCAAAAGGGTCATCATCT-3'<br>Reverse: 5′-GAGGGGCCATCCACAGTCTTCT-3' | 235 |

**Acknowledgement**

The authors declare that they have not use AI-generated work in this manuscript.

**Authors' contributions**

G.C. and D.M.Z. designed the experiments and drafted the manuscript. Y.Y.Z., Y.F., Y.Y.C., Z.Y.W., Y.C., X.C. and Y.H.J. performed the experiments. All authors contributed to the manuscript revision, read, and approved the submitted version.

**Funding**

This work was supported by the National Natural Science Foundation of China (32070998, 32271054, 31872773), Jiangsu Provincial Medical Key Discipline (Laboratory) Cultivation Unit (JSDW202249), Natural Science Foundation of Nantong City (JC2023114), and Priority Academic Program Development of Jiangsu Higher Education Institutions (PAPD).

**Availability of data and materials**

The datasets used and/or analysed during the current study are available from the corresponding author on reasonable request.All data generated or analysed during this study are included in this published article [and its supplementary information files].

**Declarations**

**Ethics approval and consent to participate**

The animal protocol was reviewed and approved by the Institutional Animal Care and Use Committee of Nantong University (Title: Role and mechanism of TLR2 protein in spinal cord NG2 cells in neuropathic pain in mice; Approval number: S20220303-005; Date: March 3, 2022). Human bone marrow mesenchymal stem cells were obtained following signed informed consent and approved the project: Human Muse Cells Induced Differentiation of neural precursor Cells and Functional Neurons and Repair of Spinal Cord Injury, the Ethics Committee of Affiliated Hospital of Nantong University (NO:2015-052). March 9, 2015.

**Consent for publication**

Not applicable.

**Competing interests**

The authors declare no competing financial interests.

# References

1. Bannister, K., Sachau, J., Baron, R. & Dickenson, A. H. Neuropathic Pain: Mechanism-Based Therapeutics. Annu Rev Pharmacol Toxicol, 2020 257-274, doi:10.1146/annurev-pharmtox-010818-021524.

2. Pittenger, M. F., Mackay, A. M., Beck, S. C., Jaiswal, R. K., Douglas, R., Mosca, J. D. et al. Multilineage Potential of Adult Human Mesenchymal Stem Cells. Science, 1999 143-147, doi:10.1126/science.284.5411.143.

3. Huh, Y., Ji, R. R. & Chen, G. Neuroinflammation, Bone Marrow Stem Cells, and Chronic Pain. Front Immunol, 2017 1014, doi:10.3389/fimmu.2017.01014.

4. Buchheit, T., Huh, Y., Maixner, W., Cheng, J. & Ji, R. R. Neuroimmune Modulation of Pain and Regenerative Pain Medicine. J Clin Invest, 2020 2164-2176, doi:10.1172/JCI134439.

5. Kuroda, Y., Kitada, M., Wakao, S. & Dezawa, M. Bone Marrow Mesenchymal Cells: How Do They Contribute to Tissue Repair and Are They Really Stem Cells? Arch Immunol Ther Ex, 2011 369-378, doi:10.1007/s00005-011-0139-9.

6. Guo, W., Imai, S., Yang, J. L., Zou, S. P., Watanabe, M., Chu, Y. X. et al. In Vivo Immune Interactions of Multipotent Stromal Cells Underlie Their Long-Lasting Pain-Relieving Effect. Sci Rep-Uk, 2017, doi:ARTN 10107 10.1038/s41598-017-10251-y.

7. Garay-Mendoza, D., Villarreal-Martinez, L., Garza-Bedolla, A., Perez-Garza, D. M., Acosta-Olivo, C., Vilchez-Cavazos, F. et al. The Effect of Intra-Articular Injection of Autologous Bone Marrow Stem Cells on Pain and Knee Function in Patients with Osteoarthritis. Int J Rheum Dis, 2018 140-147, doi:10.1111/1756-185x.13139.

8. Abbuehl, J. P., Tatarova, Z., Held, W. & Huelsken, J. Long-Term Engraftment of Primary Bone Marrow

Stromal Cells Repairs Niche Damage and Improves Hematopoietic Stem Cell Transplantation. Cell Stem Cell, 2017 241-255 e246, doi:10.1016/j.stem.2017.07.004.

9. Lu, S., Ge, M., Zheng, Y., Li, J., Feng, X., Feng, S. et al. Cd106 Is a Novel Mediator of Bone Marrow Mesenchymal Stem Cells Via Nf-Kappab in the Bone Marrow Failure of Acquired Aplastic Anemia. Stem Cell Res Ther, 2017 178, doi:10.1186/s13287-017-0620-4.

10. Wakao, S., Kushida, Y. & Dezawa, M. Basic Characteristics of Muse Cells. Adv Exp Med Biol, 2018 13-41, doi:10.1007/978-4-431-56847-6_2.

11. Kuroda, Y., Kitada, M., Wakao, S., Nishikawa, K., Tanimura, Y., Makinoshima, H. et al. Unique Multipotent Cells in Adult Human Mesenchymal Cell Populations. Proc Natl Acad Sci U S A, 2010 8639-8643, doi:10.1073/pnas.0911647107.

12. Dezawa, M. Muse Cells Provide the Pluripotency of Mesenchymal Stem Cells: Direct Contribution of Muse Cells to Tissue Regeneration. Cell Transplant, 2016 849-861, doi:10.3727/096368916x690881.

13. Iseki, M., Kushida, Y., Wakao, S., Akimoto, T., Mizuma, M., Motoi, F. et al. Human Muse Cells, Nontumorigenic Pluripotent-Like Stem Cells, Have Liver Regeneration Capacity through Specific Homing and Cell Replacement in a Mouse Model of Liver Fibrosis. Cell Transplant, 2017 821-840, doi:10.3727/096368916x693662.

14. Fei, W., Wu, J., Gao, M., Wang, Q., Zhao, Y. Y., Shan, C. et al. Multilineage-Differentiating Stress-Enduring Cells Alleviate Atopic Dermatitis-Associated Behaviors in Mice. Stem Cell Res Ther, 2021 606, doi:10.1186/s13287-021-02671-5.

15. Bennett, G. J. & Xie, Y. K. A Peripheral Mononeuropathy in Rat That Produces Disorders of Pain Sensation Like Those Seen in Man. Pain, 1988 87-107, doi:10.1016/0304-3959(88)90209-6.

16. Chen, G., Park, C. K., Xie, R. G. & Ji, R. R. Intrathecal Bone Marrow Stromal Cells Inhibit Neuropathic Pain Via Tgf-Beta Secretion. J Clin Invest, 2015 3226-3240, doi:10.1172/JCI80883.

17. Goncalves, N. P., Jager, S. E., Richner, M., Murray, S. S., Mohseni, S., Jensen, T. S. et al. Schwann Cell P75 Neurotrophin Receptor Modulates Small Fiber Degeneration in Diabetic Neuropathy. Glia, 2020 2725-2743, doi:10.1002/glia.23881.

18. Colvin, L. A. Chemotherapy-Induced Peripheral Neuropathy: Where Are We Now? Pain, 2019 S1-S10, doi:10.1097/j.pain.0000000000001540.

19. Berta, T., Qadri, Y., Tan, P. H. & Ji, R. R. Targeting Dorsal Root Ganglia and Primary Sensory Neurons for the Treatment of Chronic Pain. Expert Opin Ther Tar, 2017 695-703, doi:10.1080/14728222.2017.1328057.

20. Chen, G., Zhang, Y. Q., Qadri, Y. J., Serhan, C. N. & Ji, R. R. Microglia in Pain: Detrimental and Protective Roles in Pathogenesis and Resolution of Pain. Neuron, 2018 1292-1311, doi:10.1016/j.neuron.2018.11.009.

21. Ji, R. R., Nackley, A., Huh, Y., Terrando, N. & Maixner, W. Neuroinflammation and Central Sensitization in Chronic and Widespread Pain. Anesthesiology, 2018 343-366, doi:10.1097/ALN.0000000000002130.

22. Matsuda, M., Huh, Y. & Ji, R. R. Roles of Inflammation, Neurogenic Inflammation, and Neuroinflammation in Pain. J Anesth, 2019 131-139, doi:10.1007/s00540-018-2579-4.

23. Sasaki, M., Abe, R., Fujita, Y., Ando, S., Inokuma, D. & Shimizu, H. Mesenchymal Stem Cells Are Recruited into Wounded Skin and Contribute to Wound Repair by Transdifferentiation into Multiple Skin Cell Type. J Immunol, 2008 2581-2587, doi:10.4049/jimmunol.180.4.2581.

24. Mburu, Y. K., Wang, J., Wood, M. A., Walker, W. H. & Ferris, R. L. Ccr7 Mediates Inflammation-

Associated Tumor Progression. Immunol Res, 2006 61-72, doi:10.1385/IR:36:1:61.

25. Honczarenko, M., Le, Y., Swierkowski, M., Ghiran, I., Glodek, A. M. & Silberstein, L. E. Human Bone Marrow Stromal Cells Express a Distinct Set of Biologically Functional Chemokine Receptors. Stem Cells, 2006 1030-1041, doi:10.1634/stemcells.2005-0319.

26. Finnerup, N. B., Kuner, R. & Jensen, T. S. Neuropathic Pain: From Mechanisms to Treatment. Physiol Rev, 2021 259-301, doi:10.1152/physrev.00045.2019.

27. Bianco, P., Riminucci, M., Gronthos, S. & Robey, P. G. Bone Marrow Stromal Stem Cells: Nature, Biology, and Potential Applications. Stem Cells, 2001 180-192, doi:10.1634/stemcells.19-3-180.

28. Wakao, S., Kuroda, Y., Ogura, F., Shigemoto, T. & Dezawa, M. Regenerative Effects of Mesenchymal Stem Cells: Contribution of Muse Cells, a Novel Pluripotent Stem Cell Type That Resides in Mesenchymal Cells. Cells, 2012 1045-1060, doi:10.3390/cells1041045.

29. Chen, N. F., Chen, W. F., Sung, C. S., Lu, C. H., Chen, C. L., Hung, H. C. et al. Contributions of P38 and Erk to the Antinociceptive Effects of Tgf-Beta1 in Chronic Constriction Injury-Induced Neuropathic Rats. J Headache Pain, 2016 72, doi:10.1186/s10194-016-0665-2.

30. Luckemeyer, D. D., Xie, W., Prudente, A. S., Qualls, K. A., Tonello, R., Strong, J. A. et al. The Antinociceptive Effect of Sympathetic Block Is Mediated by Transforming Growth Factor Beta in a Mouse Model of Radiculopathy. Neurosci Bull, 2023 1363-1374, doi:10.1007/s12264-023-01062-5.

31. Milligan, E. D., Sloane, E. M., Langer, S. J., Hughes, T. S., Jekich, B. M., Frank, M. G. et al. Repeated Intrathecal Injections of Plasmid DNA Encoding Interleukin-10 Produce Prolonged Reversal of Neuropathic Pain. Pain, 2006 294-308, doi:10.1016/j.pain.2006.07.009.

32. Janssens, S., Dubois, C., Bogaert, J., Theunissen, K., Deroose, C., Desmet, W. et al. Autologous Bone

Marrow-Derived Stem-Cell Transfer in Patients with St-Segment Elevation Myocardial Infarction: Double-Blind, Randomised Controlled Trial. Lancet, 2006 113-121, doi:10.1016/S0140-6736(05)67861-0.

33. Naji, A., Eitoku, M., Favier, B., Deschaseaux, F., Rouas-Freiss, N. & Suganuma, N. Biological Functions of Mesenchymal Stem Cells and Clinical Implications. Cell Mol Life Sci, 2019 3323-3348, doi:10.1007/s00018-019-03125-1.

34. Le Blanc, K. & Mougiakakos, D. Multipotent Mesenchymal Stromal Cells and the Innate Immune System. Nat Rev Immunol, 2012 383-396, doi:10.1038/nri3209.

35. Wang, Z., Chai, C., Wang, R., Feng, Y., Huang, L., Zhang, Y. et al. Single-Cell Transcriptome Atlas of Human Mesenchymal Stem Cells Exploring Cellular Heterogeneity. Clin Transl Med, 2021 e650, doi:10.1002/ctm2.650.

36. Izadpanah, R., Kaushal, D., Kriedt, C., Tsien, F., Patel, B., Dufour, J. et al. Long-Term in Vitro Expansion Alters the Biology of Adult Mesenchymal Stem Cells. Cancer Res, 2008 4229-4238, doi:10.1158/0008-5472.CAN-07-5272.

37. Guo, W., Wang, H., Zou, S., Gu, M., Watanabe, M., Wei, F. et al. Bone Marrow Stromal Cells Produce Long-Term Pain Relief in Rat Models of Persistent Pain. Stem Cells, 2011 1294-1303, doi:10.1002/stem.667.

38. Suzuki, T., Sato, Y., Kushida, Y., Tsuji, M., Wakao, S., Ueda, K. et al. Intravenously Delivered Multilineage-Differentiating Stress Enduring Cells Dampen Excessive Glutamate Metabolism and Microglial Activation in Experimental Perinatal Hypoxic Ischemic Encephalopathy. J Cereb Blood Flow Metab, 2021 1707-1720, doi:10.1177/0271678X20972656.

39. Fisher, S. A., Brunskill, S. J., Doree, C., Mathur, A., Taggart, D. P. & Martin-Rendon, E. Stem Cell

Therapy for Chronic Ischaemic Heart Disease and Congestive Heart Failure. Cochrane Database Syst Rev, 2014 CD007888, doi:10.1002/14651858.CD007888.pub2.

40. Yeung, C. K., Yan, Y., Yan, L., Duan, Y., Li, E., Huang, B. et al. Preclinical Safety Evaluation and Tracing of Human Mesenchymal Stromal Cell Spheroids Following Intravenous Injection into Cynomolgus Monkeys. Biomaterials, 2022 121759, doi:10.1016/j.biomaterials.2022.121759.